\documentclass{SIP}

\usepackage{graphicx}
\usepackage{caption}
\usepackage{subcaption}
\usepackage{multirow}
\usepackage{hyperref} 
\usepackage{siunitx}
\usepackage{amssymb}
\usepackage[toc,page]{appendix}
\usepackage{xr}
\usepackage[table,xcdraw]{xcolor}
\usepackage{booktabs}
\usepackage{CJKutf8}
\usepackage{tabularx}
\usepackage{tabulary} 
\usepackage{float}
\usepackage{enumitem}
\usepackage{changepage}
\usepackage{epstopdf, epsfig}
\usepackage{natbib}
\usepackage[autostyle]{csquotes}
\usepackage{rotating}
\usepackage{adjustbox}

\begin{document}

\title[A network-based approach to QAnon user dynamics and topic diversity during the COVID-19 infodemic]{A network-based approach to QAnon user dynamics and topic diversity during the COVID-19 infodemic}

\author[Wentao Xu and Kazutoshi Sasahara.]{Wentao Xu$^{1}$, Kazutoshi Sasahara$^{2}$}

\address{\add{1}{Graduate School of Informatics, Nagoya University, Japan}
\add{2}{School of Environment and Society, Tokyo Institute of Technology, Japan}}

\corres{\name{Wentao Xu, Kazutoshi Sasahara}
\email{myrainbowandsky@gmail.com, sasahara.k.aa@m.titech.ac.jp}}

\begin{abstract}
QAnon is an umbrella conspiracy theory that encompasses a wide spectrum of people. The COVID-19 pandemic has helped raise the QAnon conspiracy theory to a wide-spreading movement, especially in the US. Here, we study users’ dynamics on Twitter related to the QAnon movement (i.e., pro-/anti-QAnon and less-leaning users) in the context of the COVID-19 infodemic and the topics involved using a simple network-based approach. 
We found that pro- and anti-leaning users show different population dynamics and that late less-leaning users were mostly anti-QAnon. 
These trends might have been affected by Twitter's suspension strategies.
We also found that QAnon clusters include many bot users. Furthermore, our results suggest that QAnon continues to evolve amid the infodemic and does not limit itself to its original idea but instead extends its reach to create a much larger umbrella conspiracy theory. The network-based  approach in this study is important for nowcasting the evolution of the QAnon movement.
\end{abstract}

\keywords{COVID-19, infodemic, networks, QAnon conspiracy theory, topic diversity, user dynamics}

\maketitle

\section{Introduction}
\subsection{A brief history of QAnon} 
With the worldwide rise of populism in recent years, many conspiracy theories have become increasingly popular. 
Conspiracy theories and populism are relevant to each other. 
They usually contain two roles, i.e., the powerful elites who control social resources and privilege, and the ordinary people described as the vulnerable victims \cite{Elite}. 

One of the most popular conspiracy theories is QAnon. 
QAnon is a conspiracy theory umbrella that encompasses a wide spectrum of people, including Trump supporters, COVID-19 deniers, and anti-vaxxers. 
An anonymous government official known as `Q' emerged on 4chan (anonymous English-language forum) in 2017, declaring that there was a cabal of upper hierarchy elites controlling the States, using their power to covertly abuse children (\#pizzagate); The theory encourages people to follow Donald Trump (this conspiracy theory emerged during his presidency) and believes that Trump will arrest all the members in the `Deep State' including Hillary Clinton and Barack Obama and finally bring the cabal to justice \cite{aliapoulios2021gospel, zuckerman2019qanon, 10.3389/fsoc.2020.615727}.
Although QAnon is not an extreme organization, extremists existed amongst the QAnon movement. 
On January 6\textsuperscript{th} 2021, an organized group of pro-Trump protesters rushed into the US Capitol building. This well-known violence proved that far-right extremists existed amongst QAnon are present believers.

During the COVID-19 pandemic, QAnon has used controversial and popular social topics to get more exposure. 
For instance, QAnon conspiracy theories blamed China for its long-term cover-up of the coronavirus; diffused an idea that mandated quarantine helped protect Joe Biden during the election; questioned the travel ban and advocated the use of hydroxychloroquine; arbitrarily connected COVID-19 to the presidential election and China so that the coronavirus was just a media-hyped tool to secure the Democrats' victory in the election, and even introduced a discord element such as `Black Lives Matter' to the 2020 US presidential election~\cite{Hannah_2021}.

Meanwhile, QAnon arbitrarily connected COVID-19 to the US presidential election and China to extend its beliefs~\cite{Hannah_2021}.
Surveys about the QAnon conspiracy theory discovered that the majority of the US citizens who have heard of QAnon think the conspiracy theory is harmful to the country \cite{pew}. 
There are, however, many people holding positions between the two extremes (referred to as `less-leaning users') who consider QAnon as neither harmful nor helpful; they can not be neglected as they have the potential to become the pro-QAnon in the long run.

QAnon was present on mainstream social network working services (SNSs) for a long time before Facebook, Twitter, and YouTube realized that the poor reputation of the QAnon conspiracy might induce more social problems.
QAnon followers tend to use violent rhetoric on Twitter~\cite{plancksamuel}.
In 2020, these platforms removed thousands of QAnon accounts \cite{jacksonqanon}. Facing this reality, QAnon supporters began to look for new spirit homes on SNSs, such as Parler and Telegram. Parler is a US micro-blog SNS and is famous for Trump supporters' discussions. There are active QAnon channels for QAnon discussions across various countries on Telegram~\cite{hoseini2021globalization}.
QAnon is still cloaked in mystery but one thing that is certain is the COVID-19 infodemic has helped it spread around the world.

\subsection{Related work}
The interaction between information about COVID-19 and the epidemic has shed light on the epidemiology policy and local neighbourhood's attitude towards the expert's advice \cite {Briand2021}.
The COVID-19 infodemic is a situation where the overabundance of COVID-19 related mis/disinformation is exploding on SNSs, making it difficult for people to retrieve trustful information about the pandemic.

Some research has analysed the linguistic features of the QAnon phenomenon. 
\cite{aliapoulios2021gospel} built a dataset of 4,949 `Q drops' and found that they were not generated by a single person, indicating there are apocrypha in those drops.
\cite{10.1145/3479855} analysed 483 linguistic features and designed a computational framework for analysing dissonance self-disclosures and computing the changes in user engagement surrounding dissonance. 
\cite{hoseini2021globalization} used a BERT-based topic model to examine the QAnon discourse across multiple languages and discovered that the German language is prevalent in QAnon groups and channels on Telegram.
\cite{10.1145/3463677.3463718} used VADER to assess QAnon-related users' positions towards Trump and Biden and employed a BERT model to describe user profiles. They found that the majority of QAnon users were Donald Trump supporters, and their Twitter profiles contain `MAGA', `God', `Patriot' and `WWG1WGA'. 
\cite{Miller_2021} analysed QAnon comments on YouTube and found substantial international discussions about China, Russia, and Israel. 
These findings addressing the linguistic features suggest that the QAnon conspiracy is prevalent online and that QAnon has become a worldwide presence.

Yet other research has applied networks to address semantic aspects of the QAnon conspiracy theory. \cite{PapasavvaBSZC21} identified QAnon-relevant words graphs using a word embedding on the Voat community. 
\cite{hanley_kumar_durumeric_2021} generated a QAnon-related domain network and trained a random forest classifier that classified misinformation and genuine news sites.

Nowadays, the task for SNSs to detect QAnon communities and ban malicious users is becoming more complex. 
It was not until January 2021 that Twitter's rules and policies gained considerable public  attention. 
It was reported that 355K Twitter users involved in the controversy over the 2020 US Presidential Election had been removed \cite{DBLP:journals/corr/abs-2101-09575}. 
In addition, Twitter removed more than 70,000 accounts that diffused harmful QAnon-associated content after the well-known US Capitol riots in January 2021 \cite{twitter}. 
\cite{Chowdhury2020OnTP} has discovered that more than 60\% of the purged users survived for more than two years before they were removed by Twitter, which questions whether the purge was efficient enough. 
Meanwhile, whether or not the removal of misbehaving users contributes to a healthier social community is still controversial, especially for QAnon users. 

\subsection{Research questions}
QAnon appears to take the advantage of the overabundance of COVID-19 mis/disinformation to gain political influence. It spreads mis/dis-information and induces negative emotions, which are harmful to `less-leaning users'---those who do not have a special preference for QAnon but have the potential to become pro-QAnon in the long run. 
Although several aspects of QAnon have been investigated as mentioned above, there is a lack of evidence as to how QAnon evolved during the COVID-19 infodemic in terms of user dynamics and topic diversity.

Our research questions are summarized as follows and we will address them using a simple network-based approach:

\hfill \break
\textbf{RQ1}: What is the pro- and anti-QAnon user dynamics during the COVID-19 infodemic?

\hfill \break
\textbf{RQ2}: What kind of topics do QAnon users spread during the COVID-19 infodemic?

\section{Data and Methods}
In this section, we explain our dataset and methods used for a network-based approach to characterize QAnon dynamics during the COVID-19 infodemic.  

\subsection{Data}
Over a 12 months period between February 20 2020 and March 1, 2021 we used the Twitter Search API to collect 880,278,195 posts from 58,519,206 unique users (including tweets and retweets) by querying COVID-19-related keywords: `corona virus', `coronavirus', `covid19', `2019-nCoV', `SARS-CoV-2', `wuhanpneumonia.' This dataset is named the \textbf{base dataset}.
In addition, we filtered English language tweets containing at least one of the terms `QAnon', `\#QAnon' or `deep state', producing 308,631 tweets from 135,740 accounts \footnote{\url{https://github.com/myrainbowandsky/A-network-based-approach-to-QAnon-user-dynamics-during-COVID-19-infodemic}}. This subset is named the \textbf{QAnon dataset}.
Both datasets were used in this study.

\subsection{Identification of pro-/anti-QAnon users and their leaning}
As QAnon is a conspiracy theory which has triggered opinions both for and against its claims, we expected to identify a characteristic retweet (RT) network where pro- and anti-users are segregated.
We constructed an RT network using the QAnon dataset and applied the $k$-core decomposition ($k=2$) \cite{6137224} to identify pro- and anti-QAnon users, where each node represented a user and directed edges between nodes represented retweets.
As expected, this resulted in an RT network with two major clusters. We decided which cluster corresponded to the pro- or anti-QAnon group by manually examining large indegree users in each cluster (who were retweeted many times) in terms of their tweets and profile descriptions.

To confirm whether the classification of pro- and anti-QAnon users was reliable enough, we conducted a manual verification as follows.
We conducted the manual verification by dividing all users into two classes. 
Two coders participated in this task and classified 60 randomly selected accounts, with 30 labeled as pro-QAnon and the other 30 labeled as anti-QAnon. 
Providing them with these account names, we asked them to read the profiles and tweets of each user and classify them into pro-QAnon and anti-QAnon. 
Then, we checked the consistency of their classifications by computing Cohen's kappa. 
The resulting kappa was 0.76, which indicated substantial agreement and certified our user classification result as statistically reliable~\footnote{Note that according to~\cite{10.2307/2529310}, Cohen's kappa value is interpreted as follows: 0.0-0.2 for slight agreement; 0.2-0.4 for fair agreement; 0.4-0.6 for moderate agreement; 0.6-0.8 for substantial agreement; and 0.8-1.0 for near perfect agreement.}.

Additionally, we defined `QAnon-leaning' as follows and identified three types of users:
`pro-leaning users', `anti-leaning users', and `less-leaning users'. 

\begin{equation}
\label{eq1}
L=\frac{P-A}{P+A}, L \in [-1,1]
\end{equation}
, where $P$ is the number of retweets from pro-QAnon users and $A$ is the number of retweets from anti-QAnon users. 
$L$ compares the leaning of a user between pro-QAnon and anti-QAnon based on retweet tendencies.
If a user has more than 70\% probability to retweet from the pro-QAnon side, this user is considered pro-leaning, and vice versa.
Thus, $-0.4 \leq L \leq 0.4$  indicates that the user is less-leaning; 
$L > 0.4$ indicate that the user is pro-leaning; 
$L < -0.4$ indicates that the user is anti-leaning.

Please remember that QAnon-leaning is quantified by $L$ (Eq.\ref{eq1}), whereas pro- and anti-QAnon classifications are based on a retweet network clustering, whose result was validated as mentioned above. 
With these, we characterised a transient dynamics of pro- and anti-QAnon users in relation to QAnon-leaning.

\subsection{Human/bot classification}
To classify users into bots and humans, we used the Botometer API v4.
The Botometer is a well-recognized tool for automatically detecting bots based on supervised machine learning.
The Botometer model is trained with 1200 features, covering six categories including the account's metadata,  retweet and mention networks, temporal features, content information and sentiment \cite{Sayyadiharikandeh_2020}.
The Botometer has been applied in a series of studies to quantify the online behaviours of bots~\cite{Shao2018,Vosoughi1146}. 
This tool computes `complete automation probability' (CAP) for each user that ranges within the range of $[0,1]$; The higher the value, the higher the probability that the user is a bot. 
In this study, we set CAP=$0.7$ as the threshold for human/bot classification, which means if the CAP for a user is larger than 0.7, this user is considered to be a bot.
We validated that this threshold provides a reliable human/bot classification in a previous study using the same dataset~\cite{Xu2021}. 

\subsection{Hashtag co-occurrence networks for topics}
Latent Dirichlet Allocation (LDA) is a standard method for modelling topics from a given text~\cite{10.5555/944919.944937}.
However, LDA often fails to extract clear topics from tweets, as their text length is too short. 
In actuality, we applied LDA modelling using the library pyLDAvis \cite{sievert-shirley-2014-ldavis} with retweets but did not obtain insightful topics (See Table~\ref{tab:LDA} in the Appendix).

Therefore, we determined to use a hashtag co-occurrence network to observe topic diversity for the QAnon conspiracy theory.
This network is simple but useful for looking at complex relationships among topics, which can not be achieved by LDA-extracted topics.
We constructed hashtag co-occurrence networks for both the base and QAnon datasets in order to understand the topical diversity of the QAnon conspiracy theory.
In this network, each node is a hashtag and undirected edges between nodes represent the co-occurrence of two hashtags. 
We generated a hashtag co-occurrence network from the base dataset, applied the $k$-core decomposition ($k=10$) to it, and then extracted all the neighbours of `\#QAnon' and itself. 
Recall that the base dataset includes multiple languages (not only English). 
From the resulting network, we generated a hashtag co-occurrence network ($1000$-core) for further analysis and obtained 336 unique hashtags (nodes).
Similarly, we constructed a hashtag co-occurrence network ($k=10$-core) from the QAnon dataset that contains only English tweets and obtained 323 unique hashtags.

The modularity-based community detection algorithm, the Louvain method~\cite{blondel2008fast}, was applied to the hashtag co-occurrence network to identify clusters using the Gephi software ~\cite{Jacomy2014} \footnote{\url{https://gephi.org/}}. 
Finally, we assigned the resulting modularity class IDs to each node of the hashtag co-occurrence network for further analyses. 

\subsection{Hashtag semantic map}
Next, we used an embedding technique to visualize a semantic map of QAnon hashtags. 
To this end, we extracted the top 50 degree hashtags, except most-QAnon-related hashtags, including: `\#QANON', 
`\#QANONAS', 
`\#Q',
`\#QANON2020',
`\#THESTORM', 
`\#WWG1GWA' and 
`\#WWG1WGA', 
because these hashtags could be related to any semantic clusters in the QAnon dataset and finally form a dense and giant semantic cluster.

For clustering the similar topics represented by hashtags, we trained the Word2Vec model
using the topic modelling library Gensim~\footnote{https://github.com/RaRe-Technologies/gensim} by exploiting tweet texts and hashtags. 
Then, we applied the clustering algorithm HDBSCAN\footnote{https://github.com/scikit-learn-contrib/hdbscan}
~\cite{hdbscan} to reduce the dimensionality of the Word2Vec hashtag embeddings ($d=2$) and visualised the results using UMAP\footnote{https://github.com/lmcinnes/umap} \cite{umap}.

\section{Results}
\subsection{QAnon user dynamics}
\begin{figure}[t]
    \centering
    \begin{subfigure}[b]{0.475\linewidth}
        \centering
        \includegraphics[width=\linewidth]{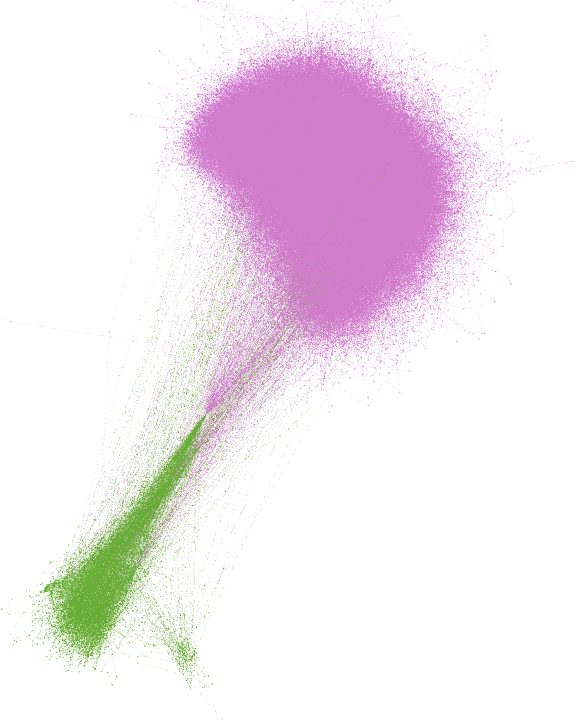}
        \caption{}\label{fig:fig1a}
    \end{subfigure}%
    \hfill
    \begin{subfigure}[b]{0.475\linewidth}  
        \centering 
        \includegraphics[width=\linewidth]{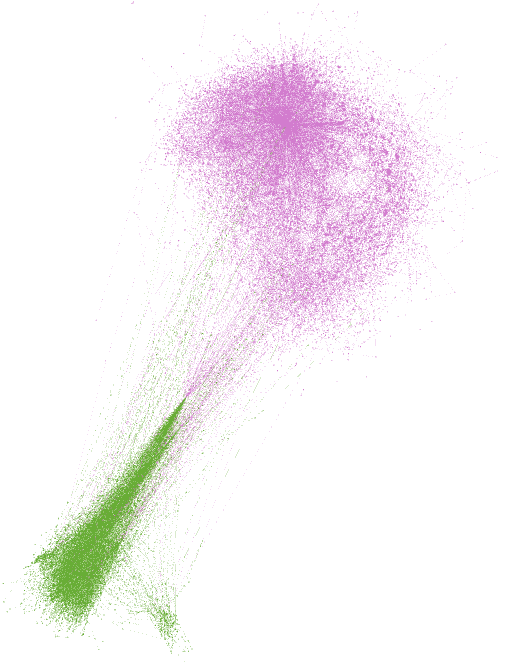}
        \caption{}\label{fig:fig1b}
    \end{subfigure}
    \hfill
\caption{Retweet network of pro-/anti-QAnon users. (a) active users from February 2020 to March 2021; (b) active users in August 2021, in which magenta denotes pro-QAnon and green denotes anti-QAnon.}  
\label{fig:fig1}
\end{figure}

Fig.~\ref{fig:fig1a} shows the retweet (RT) network ($2$-core) constructed from the QAnon dataset between February 2020 and March 2021, revealing that pro- and anti-QAnon clusters are actually segregated.
The pro-QAnon cluster ($n=40,512$) was much larger in size than the anti-QAnon cluster ($n=5,480$) (See Table~\ref{tab:alluserdemographics}).
We used these pro- and anti-QAnon classifications (validated as mentioned above) for the succeeding analyses.
We checked users' activity in August 2021 to estimate how many pro- and anti-QAnon users were suspended by Twitter.
From Fig.~\ref{fig:fig1a} to Fig.~\ref{fig:fig1b}, more than 50\% (25,318) of the users were suspended or had their accounts closed in the pro-QAnon cluster, but only 653 users faced punishment in the anti-QAnon cluster (Table \ref{tab:alluserdemographics}).

\begin{figure}[t]
  \includegraphics[width=\linewidth]{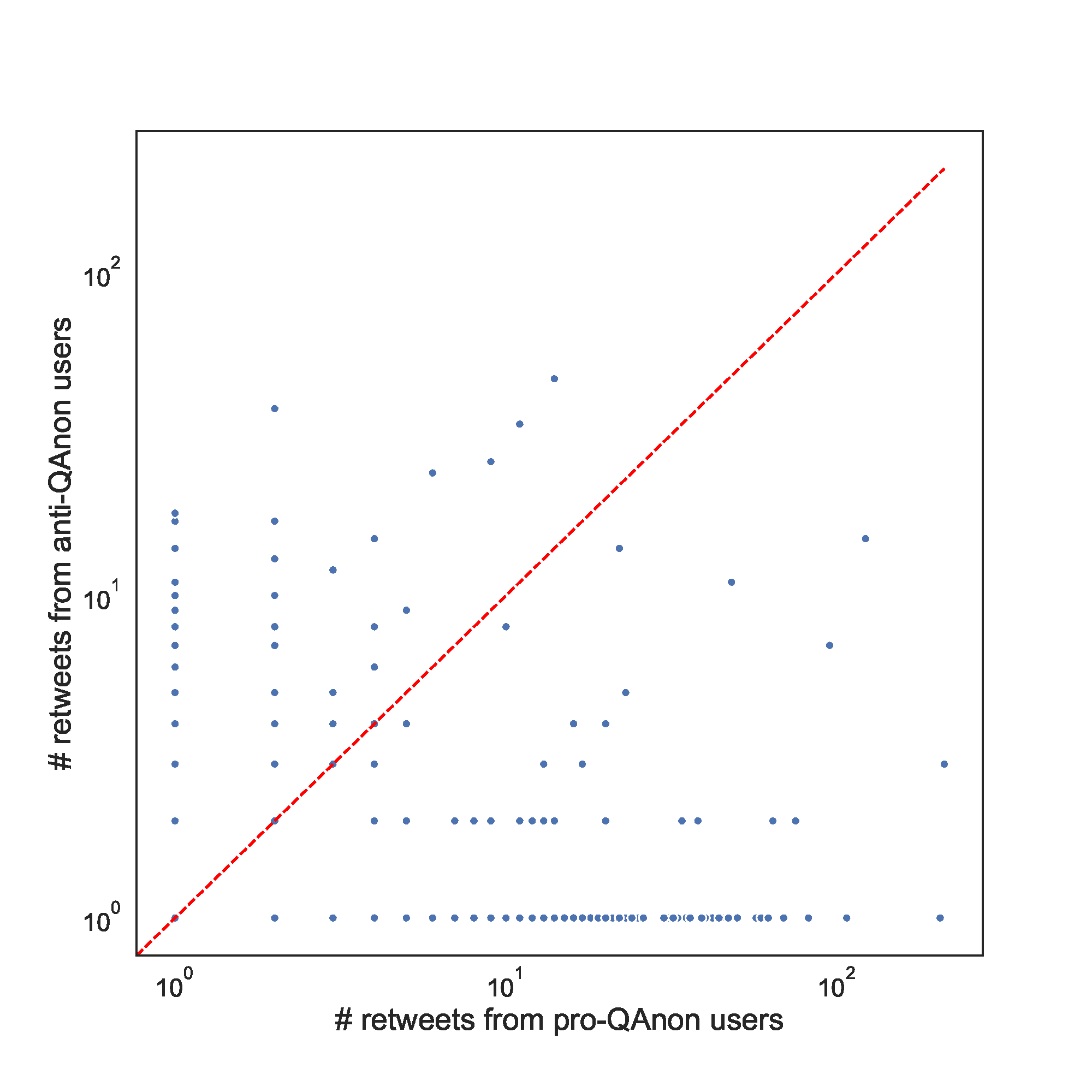}
\caption{User scatter plot with the number of retweets from pro-QAnon users and the number of retweets from anti-QAnon users (logarithmic scale).}
\label{fig:re_r}
\end{figure}

We then looked into user dynamics depending on `QAnon-leaning' ($L$).
Fig.~\ref{fig:re_r} is a user scatter plot made from the QAnon dataset, showing the relationship between the number of retweets from pro-QAnon users and those from anti-QAnon users.
This reveals that there exist not only users who retweeted most from pro-QAnon users (i.e., `pro-leaning') but also users with`anti-leaning' and `less-leaning (See Fig.~\ref{fig:violinplots}a in the Appendix).

\begin{figure}[t]
  \includegraphics[width=\linewidth]{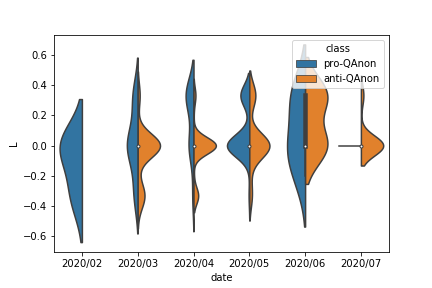}
\caption{QAnon-leaning (L) distributions for less-leaning users. Note that pro- and anti-QAnon  classifications are based on the retweet network clustering.}
\label{fig:violin}
\end{figure}

Fig.~\ref{fig:violin} shows temporal changes of QAnon-leaning ($L$) distributions for less-leaning users. 
The majority of less-leaning users are consistently centered around $0.0$ across months, except for a bi-modal peak (around 0.3) in July 2020. 
However, user types continued to gradually change. The users  in February 2020 were all pro-QAnon users but after that anti-QAnon users increased and took over pro-QAnon users in the succeeding months.

The same plots for pro- and anti-leaning users are shown in Figs.~\ref{fig:violinplots}b and c in the Appendix.
Unlike less-leaning users QAnon-leaning distributions were steadier, suggesting that both pro- and anti-leaning users were consistent in retweeted information across time.
This result suggests that Twitter's intervention by removing malicious users might have helped prevent less-leaning users from changing their preferences towards pro-QAnon.
Although less-leaning users are a minority, how to protect them from an overwhelming number of pro-QAnon group users is an urgent problem for an SNS platform like Twitter.  

\begin{figure}[H]
  \includegraphics[width=\linewidth]{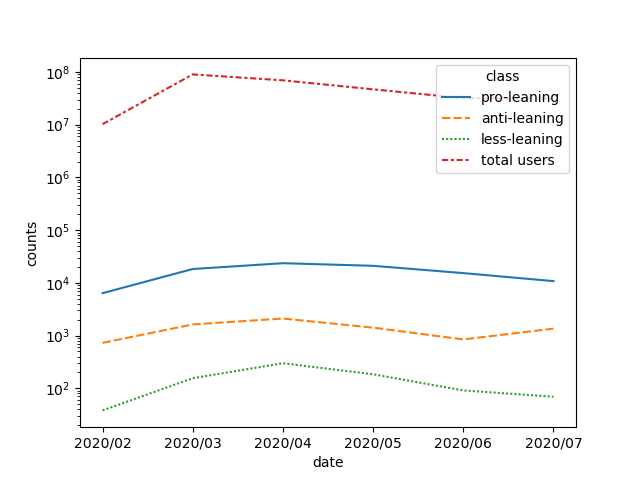}
\caption{User activity: \# of active users who retweeted at least once a month, including pro-leaning, anti-leaning, and less-leaning users, and total.}
\label{fig:user_retweet}
\medskip
\end{figure}

Then, we quantified monthly changes of active users---pro-leaning, anti-leaning, less-leaning, and total---in Fig.~\ref{fig:user_retweet}).
The total amount of active users was peaked in March 2020 and then decreased. 
However, the numbers of pro-leaning, anti-leaning, and less-leaning all peaked one month later. 
After that, the amount of pro-leaning users decreased monotonically (same for less-leaning users) whereas their anti-leaning counterpart again increased in July 2020.
Similar patterns were observed in the retweet activities of these users (See Fig.~\ref{fig:dynamic_retweets} in the Appendix).

All these results indicate that the removal of malicious users by Twitter might have contributed to some extent to reducing pro-QAnon users and increasing the anti-QAnon users.

\begin{table*}[!t]
\caption{Summary of pro- and anti-QAnon users (February 2020 to March 2021; suspended or closed accounts as of August 2021)}
\centering
\label{tab:alldemographics}
    \label{tab:alluserdemographics}
    \begin{tabular}{lll}\toprule
                              & \#pro-QAnon & \#anti-QAnon \\ \hline
        All users            & 40,512      & 5,480        \\ \hline
        Suspended users & 25,318      & 653          \\ \hline
        Bots                  & 8,239       & 2,861        \\ \hline
        Humans                & 6,016       & 2,592        \\ \bottomrule
    \end{tabular}
    
    
\end{table*}

\subsection{Prevalence of bots in QAnon clusters} 
We also examined how many bots were involved in pro- and anti-QAnon users. 
There were 8,239 bots and 6,016 humans in the pro-QAnon cluster while there were 2,861 bots and 1,592 humans in the anti-QAnon cluster. (Shown in Table \ref{tab:alluserdemographics}.
Thus bots are prevalent not only in the pro-QAnon cluster but also in the anti-QAnon cluster, playing a major role in spreading QAnon conspiracy topics, on one hand, and passing on information debunking them, on the other.
This result was different from other mis/disinformation phenomena during the COVID-19 infodemic (e.g., see ~\cite{Xu2021}).
Note that we could not obtain all the bot scores because of user suspensions by Twitter or their inaccessibility due to private settings; thus, the number of bots and humans reported here could be smaller than the actual count. 

\subsection{Hashtag co-occurrence network as a conspiracy theory umbrella}
\begin{figure*}[t]
  \includegraphics[width=\linewidth]{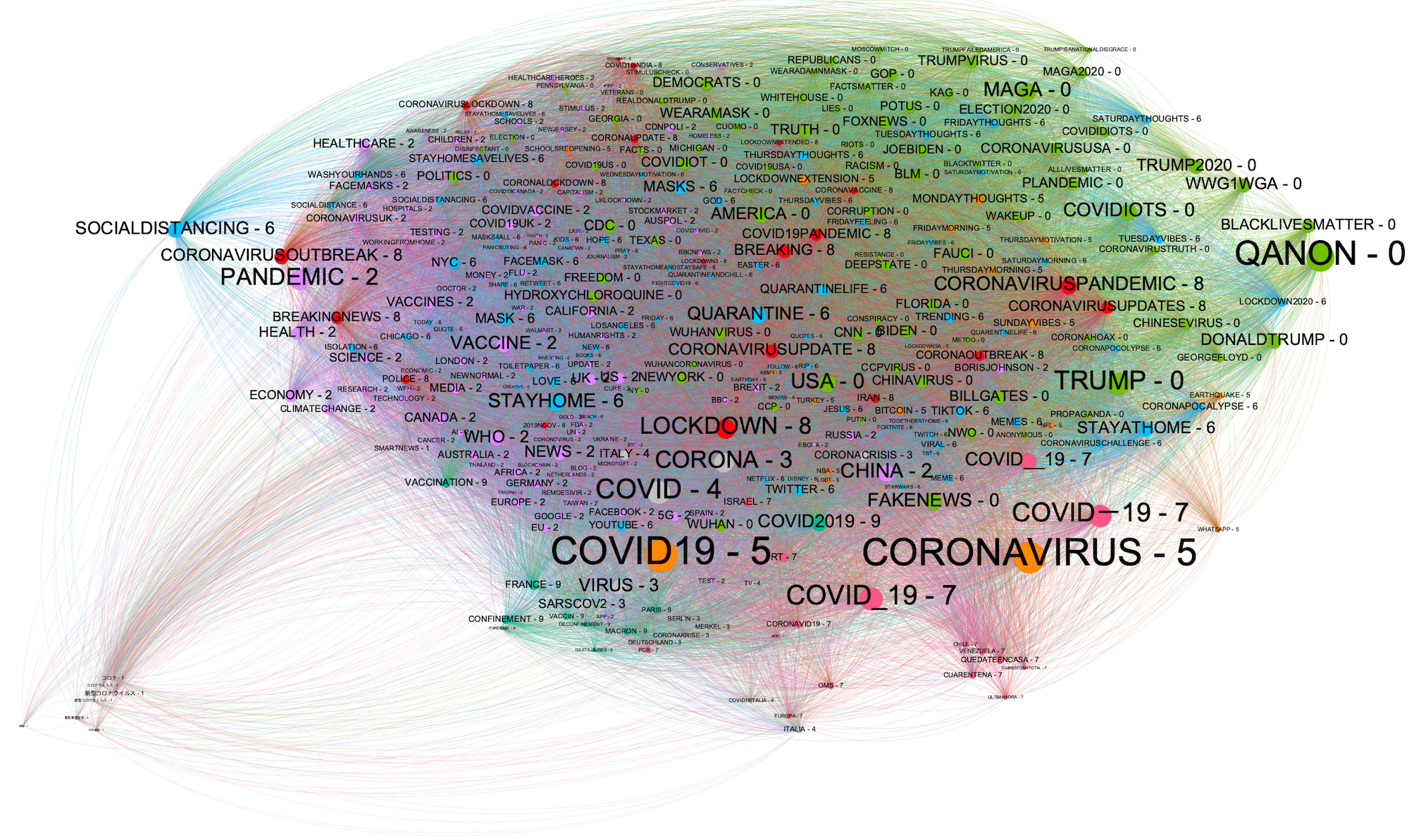}
\caption{Global hashtag co-occurrence network. Numbers denote hashtag classes. `\#QANON' is in class 0 (green). The degree is represented along with each hashtag. The label size of a node is proportional to its degree.}
\label{fig:QAnon_neighbor_hashtagNT}      
\end{figure*}

The global hashtag co-occurrence network ($1000$-core) was constructed using the base dataset.
The resulting network is illustrated in Fig.~\ref{fig:QAnon_neighbor_hashtagNT} ($n=336$).
This visualises a topic landscape for QAnon conspiracy theory in the context of the COVID-19 infodemic, as the base dataset includes multiple languages and diverse COVID-19 topics.
Here, we see that the three most popular topics are `US politics', `News' and `Daily life'. Furthermore, \#QAnon has even co-occurred with human rights hashtags, such as `\#LGBT' ($k=1,418$), `\#METOO' ($k=1,073$) and `\#BLACKLIVESMATTER' ($k=6,390$), which is consistent with \cite{Hannah_2021}. Note that $k$ denotes degree or the number of occurrences. The co-occurrence patterns of popular hashtags can reflect the topical diversity of QAnon conspiracy theory, which consequently facilitates greater exposures to users amidst the pandemic.

Furthermore, we can see an isolated cluster (class 1) of Japanese hashtags at the bottom-left of Fig.~\ref{fig:QAnon_neighbor_hashtagNT}, which is related to J-Anon, QAnon's Japanese counterpart. JAnon users also believe that (former) President Trump is a hero in the fight against the ``Deep State''.
We also find topical relations between QAnon and France (French language tweets, class 9), Spain (Spanish language tweets, class 7) , and Italy(Italian language tweets, class 4) topics, which proves that QAnon is becoming a global conspiracy theory topic, especially in Western countries. This finding supports the previous research, which suggested that the QAnon conspiracy theory originated from local niche communities including 4chan and 8chan, and then migrated to become a globally influential conspiracy theory \cite{aliapoulios2021gospel}.
In addition, the religious hashtags relevant to the `apocalypse' that Trump supporters believe in were connected to \#QAnon. 
They believed that Trump was sent by God \cite{armor_pf_god}. 
In actuality, there is a tweet mentioning `Armor of God ! ! \space \#qanon \space \#wearethenewsnow \space \#factsmatter \space \#wwg1wga \space \#wakeupamerica \space \#covid-19 \space \#unitednot'.

Because QAnon is a US conspiracy topic, we then focused on English tweets using the QAnon dataset. The 10-core English hashtag co-occurrence network ($n=232$) comprises the four conspiracy theory-related topics, including `\#WHO', `\#TRUMP', `\#5G' and `\#BILLGATES', which have been discussed previously \cite{Xu2021}. 
In addition, we observed the well-known QAnon hashtags such as `\#WWG1WGA' ($k=624$), `\#MAGA' ($k=337$), `\#THEGREATAWAKENING  ($k=244$); it seems that QAnon debunking information was also present in the network, for instance, `\#FAKENEWS' ($k=94$), `\#FAKENEWSMEDIA' ($k=15$), and `\#CONSPIRACY' ($k=31$) were identified as well. 
Since `\#FAKENEWS' is identified in both global and English hashtag co-occurrence networks, we suppose that there could be, at least, two voices towards QAnon, one is pro-QAnon and the other is anti-QAnon, which is consistent with our QAnon users' visualisations (Fig.~\ref{fig:fig1}).
In addition, we are able to identify `\#FAKENEWS' and its 64 neighbors, indicating there was a voice of debunking QAnon-related news.

To understand the topics in Fig. \ref{fig:QAnon_neighbor_hashtagNT} in detail, we examined the top-50-degree hashtags in relation to the pro- and anti-QAnon users. (See the statistical summary in Table~\ref{tab:swingusershashtags}.) The three most popular topics are the same as the ones described above: US politics (class 5), COVID-19 (class 0) and News (class 2).
These two networks indicate that QAnon has been evolving into a much larger conspiracy umbrella worldwide, which may potentially attract vulnerable users, including less-leaning users who are neutral to pro- and anti-QAnon groups.

\begin{figure*}[t]
  \includegraphics[width=\linewidth]{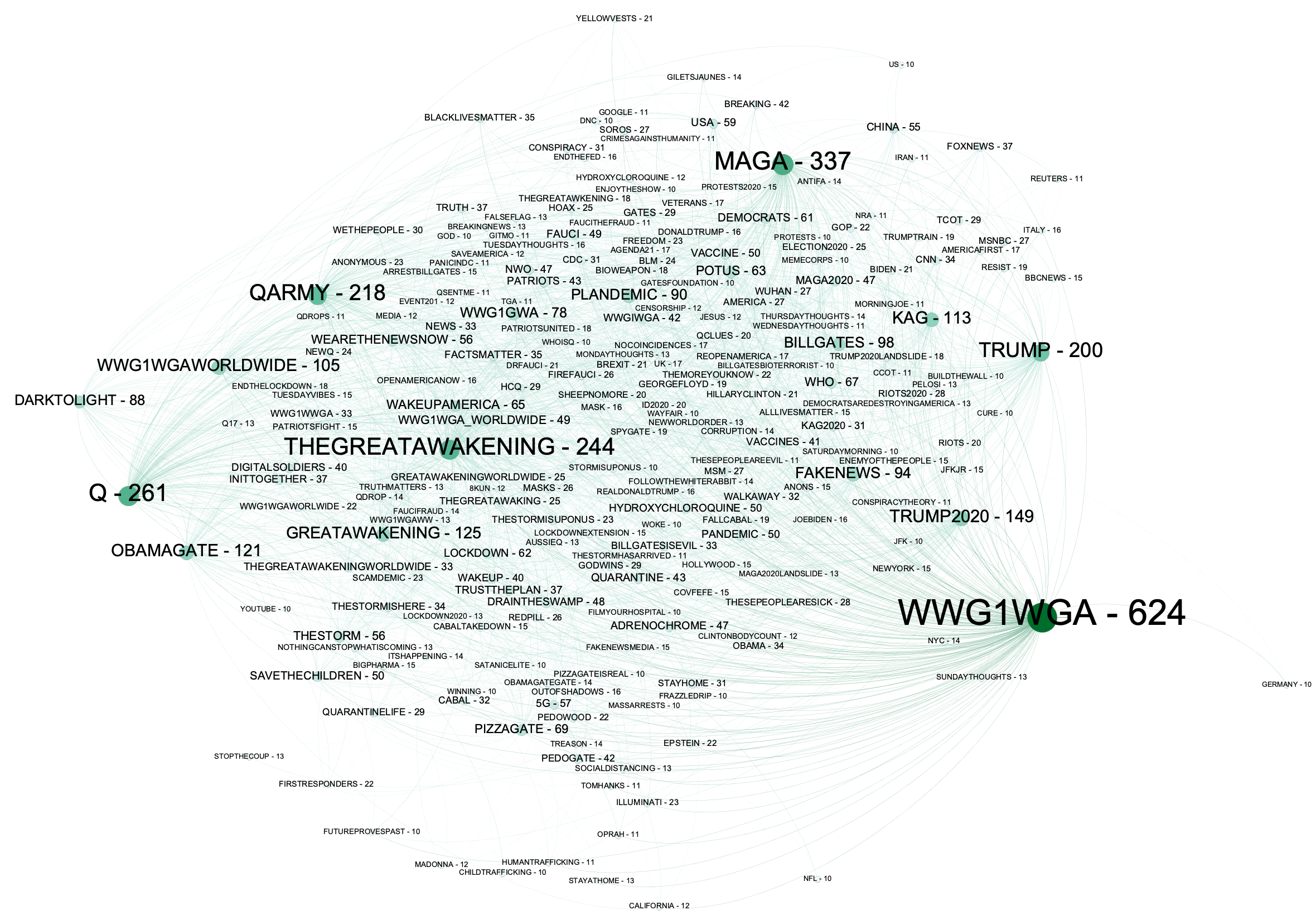}
\caption{English QAnon hashtag co-occurrence network}
\label{fig:QAnonEnglish}
\medskip
\end{figure*}

We then computed the ratio ($\%Pro/\%Anti$) of pro-users' \%hashtags to anti-users' \%hashtags to show the hashtag preference.
Here, a higher ratio means a tendency to lean towards the anti-QAnon side. If {$\%Pro/\%Anti$ > 1}, the users are holding pro-QAnon tendency in that hashtag topic; if {$\%Pro/\%Anti $ < 1}, the users are holding an anti-QAnon tendency in the topic; and if {$\%Pro/\%Anti$ = 1}, the users are holding balanced or neutral views towards the topic (Table~\ref{tab:swingusershashtags}).
Except for J-Anon, most of the hashtags showed an pro-QAnon tendency; in particular, users concerned with US politics and French tweets are more pro-QAnon.

\subsection{Hashtag semantics and dynamics}
\begin{figure*}[!t]
\includegraphics[width=\linewidth]{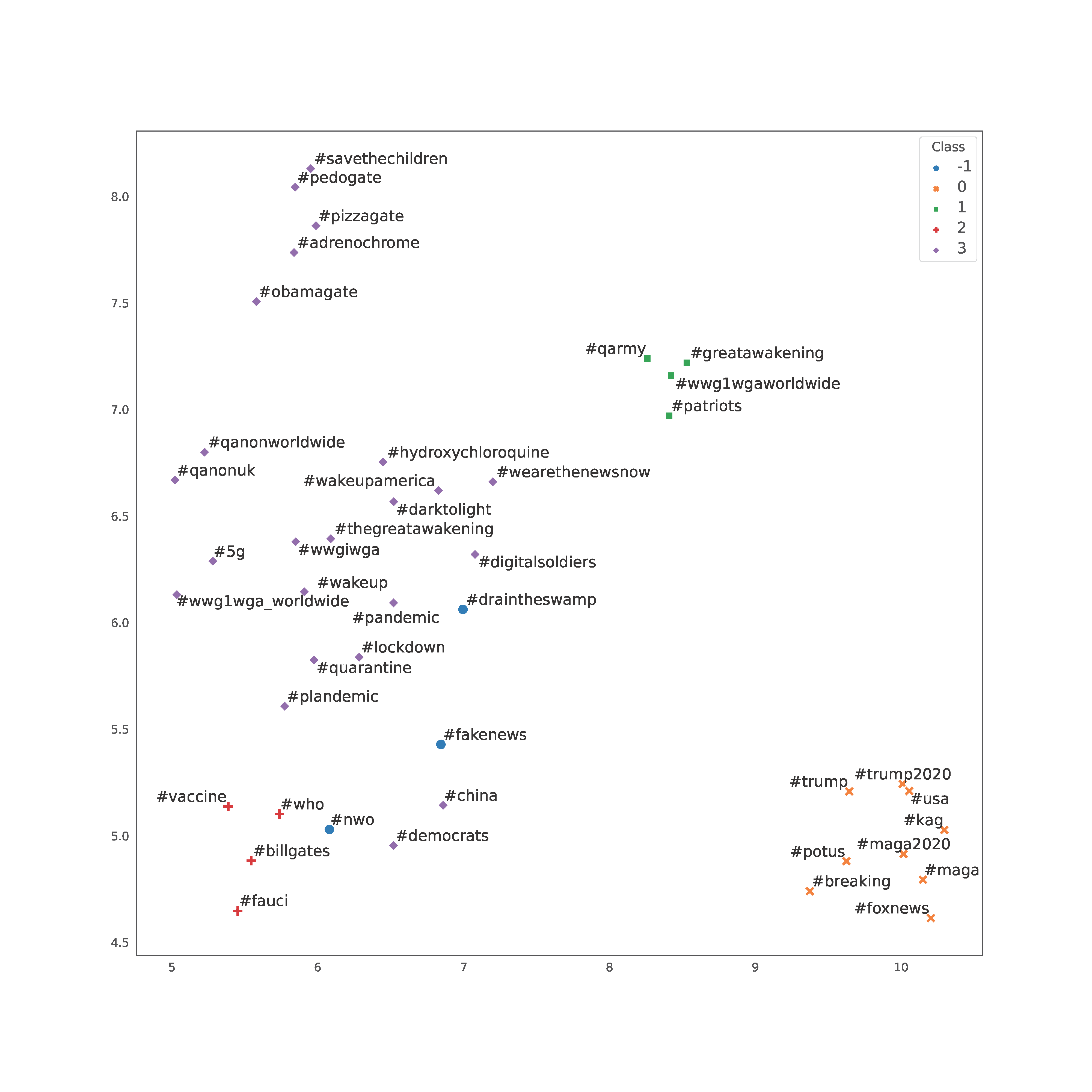}
\caption{Semantic map of top 50 popular hashtags.}
\label{fig:semanticmap}
\end{figure*}

Furthermore, we created a semantic map of the top 50 popular hashtags in the global hashtag co-occurrence network (Fig. ~\ref{fig:semanticmap}). 
Overall, semantically similar hashtags are grouped together on this map: the conservative cluster (cluster 0: e.g., \#trump, \#maga), the QAnon cluster (cluster 1: e.g., \#plandemic, \#5g), the vaccine cluster (cluster 2: e.g., \#vaccine, \#fauci ) and outliers (cluster -1: e.g., \#china, \#fakenews).
and includes diverse QAnon topics, such as \#plandemic, \#5g, \#pizzagate, and \#obamagate.
The  lexical resemblance in cluster 3 could be explained by the fact that ``\#plademic'' and QAnon were associated in the context of community victimization~\cite{10.3389/fpubh.2021.649930}. 

\begin{figure*}[p]
  \includegraphics[width=\linewidth]{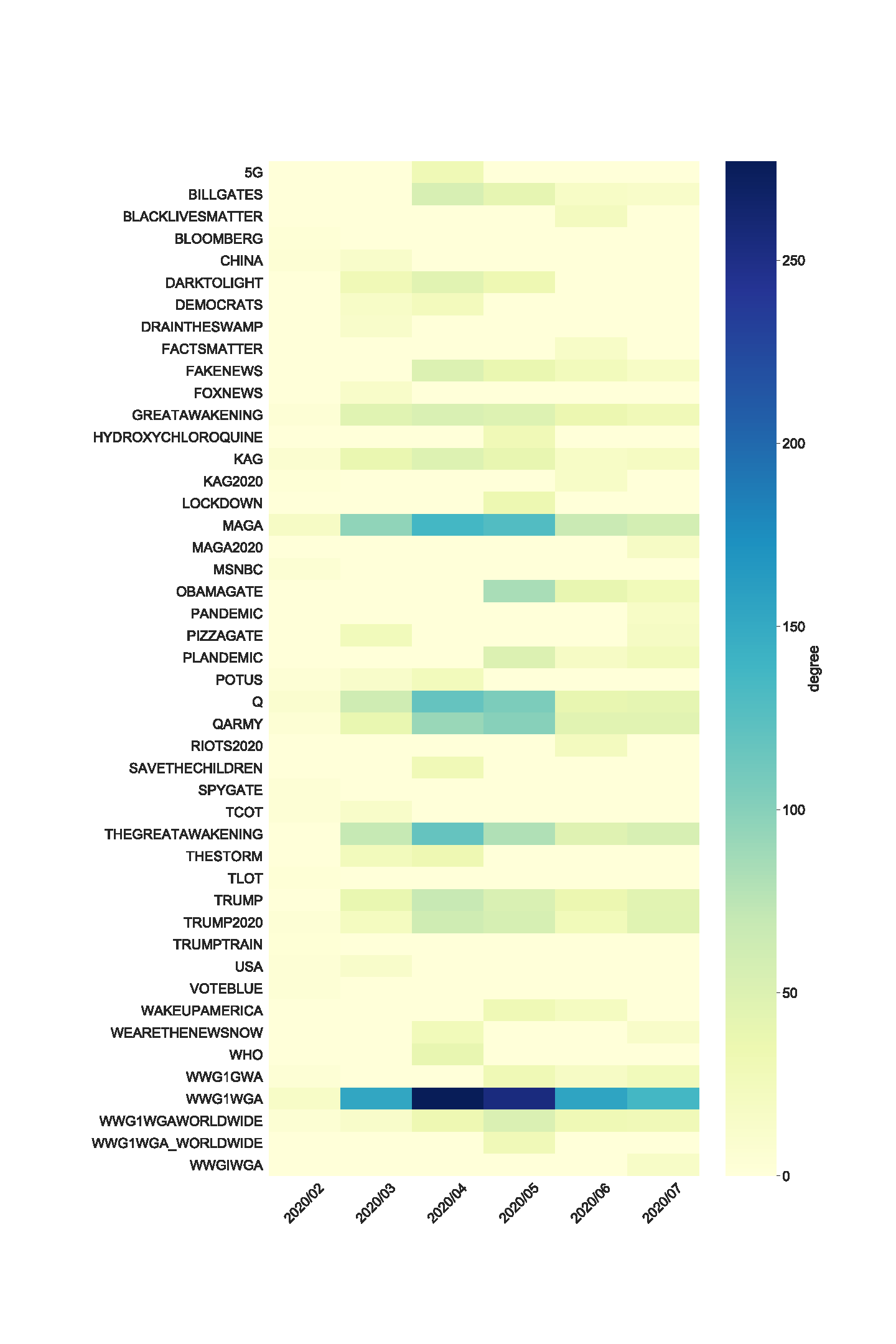}
\caption{Temporal changes of monthly top 20 popular hashtags. A darker hashtag indicates a higher degree.}
\label{fig:heatmap}       
\end{figure*}

In addition, we examined the temporal changes of hashtags occurrences (Fig.~\ref{fig:heatmap}). We found that QAnon representative hashtags, including `\#WWG1WGA', `\#Q', `\#QARMY' and `\#THEGREATAWAKENIN', appeared together in sync.
It turns out that these hashtags were involved in the gigantic component of the global hashtag co-occurrence network. 
The degrees of these hashtags reached their peaks between April and May 2020, during which QAnon's topics flourished.

\begin{table*}[!t]
\centering
\caption{Top 10 hashtags preferred by pro- and anti-QAnon users}
\label{tab:swingusershashtags}
\begin{tabular}{llll}
\toprule
Topic class & \%Pro & \%Anti & \%Pro/\%Anti \\ \hline
US politics & 80    & 20     & 4.0          \\ \hline
J-Anon      & 32    & 68     & 0.5          \\ \hline
News        & 70    & 30     & 2.3          \\ \hline
Lockdown    & 67    & 33     & 2.0          \\ \hline
Italy       & 67    & 33     & 2.0          \\ \hline
COVID-19    & 61    & 39     & 1.6          \\ \hline
Daily life  & 73    & 27     & 2.7          \\ \hline
Spain       & 72    & 28     & 2.6          \\ \hline
India       & 70    & 30     & 2.3          \\ \hline
France      & 78    & 22     & 3.5         \\\bottomrule
\end{tabular}
\end{table*}

\section{Discussion}
We summarise our results obtained from a simple network-based approach and discuss their imprecations to counter the QAnon movement. 

Regarding RQ1, we found that the pro-QAnon cluster is much larger in size than the anti-QAnon cluster even though more than 50\% of its users were suspended.
A notable finding is that the numbers of pro- and anti-leaning users were both peaked in April 2020, but then pro-leaning users monotonically decreased whereas anti-leaning users increased again in July 2020.
Furthermore, late less-leaning users were mostly anti-QAnon users.
These results suggest that Twitter's suspension strategy might have helped diminish the QAnon movement. 

However, we also think that simply removing malicious users may not have done enough to effectively combat pro-QAnon users and protect other users from the attraction of diverse pro-QAnon content.
Identifying `malicious users' is often difficult. For example, a QAnon debunker may retweet or share a pro-QAnons' posts to oppose them, and algorithms might mistakenly flag them as non-credible if only the contents are viewed.  
An alternative approach is to intervene with anti- and less-leaning users by showing trustful information sources with adequate timing to increase their activity while purging extremely pro-leaning users. 
If we can better communicate with a similar emotional tone and objective stance, less-leaning users are more likely to alter their attitudes towards the anti-QAnon side.

Regarding RQ2, we found that QAnon has been evolving into a diverse and global conspiracy theory umbrella.
Previous work \cite{amarasingam2020qanon} pointed out that QAnon lacks both a clear organisational structure and a centralization of interpretive duties, compared with other extremist organizations.
However, QAnon became a popular conspiracy theory during the COVID-19 infodemic. Not only do we find `US' featured, but QAnon also has spread to other countries including France, Spain, Italy, and Japan (JAnon). 
In addition, we can identify human rights topics, such as `\#LGBT' and `\#BLACKLIVESMATTER', as well as the COVID-19 related topics, such as `\#STAYHOME' and `\#SOCIALDISTANCING'. 
These results suggested that QAnon has been growing in a semantic network, thereby forming a larger conspiracy theory umbrella. 


We stress that neutral users, including `less-leaning users' and `a silent majority', may play a key role in the evolution of QAnon conspiracy theory. 
How to protect them from an overwhelming number of pro-QAnon group is critical for SNS platforms. 
To this end, we should better inform neutral users about the nature of QAnon to avoid the `backfire effect' of their further approaching the pro-QAnon community.
As shown, the number of pro-leaning users has been decreasing at least on the Twitter platform but they were still the majority in the later stages.
In addition, some of them might have moved to other social media platforms and are looking for a chance to revive, while increasing topical diversity to attract less-leaning users.

A network-based approach in this study provides a simple but practical tool for nowcasting the evolution of the QAnon movement in terms of social and topical dynamics.
Results based on this approach can be useful information and insights for journalists, fact-checkers, and platforms to develop effective countermeasures to QAnon movement.

\section*{Acknowledgement}
We would like thank the members of CREST projects (JPMJCR20D3 and JPMJCR17A4) for fruitful discussions.

\section*{Financial Support}
This work was supported by JST, CREST Grant Number JPMJCR20D3, Japan.


\section*{Statement of interest}
None.

\bibliographystyle{ieeetr}
\bibliography{MAIN}









\vskip2pc

\noindent \large \textbf{Biographies}

\vskip2pc

\noindent\normalsize\textbf{Wentao Xu} received his Master of Research in Regenerative Medicine from University of Bath in the UK in 2007 and received his Master of Industrial Economics from Royal Institute of Technology in Sweden in 2010. He was working as data analyst and AI engineer. He is a Ph.D. candidate of Graduate School of Informatics, Nagoya University. His main research interest is machine learning, NLP, deep learning and network science.

\vskip2pc

\noindent\textbf{Kazutoshi Sasahara} received his Ph.D. from The University of Tokyo in 2005. He worked as a researcher at several institutions including RIKEN, The University of Tokyo, and UCLA. From 2012 to 2020, he was an Assistant Professor in Graduate School of Informatics, Nagoya University. Since 2020, he is an Associate Professor in School of Environment and Society, Tokyo Institute of Technology. His research interest is computational social science and social innovation.

\begin{appendices}
The top three most popular topics are `US politics', `News' and `Daily life', described as follows.
\subsubsection{class 0, $n=84$, green : US politics}
In addition to QAnon hashtags, such as `MAGA' (Make America Great Again), `WWG1WGA' (Where We Go One We Go All), and `WAKEUP', political celebrities including `TRUMP', `BILLGATES', `JOEBIDEN' existed as well identified in the class. 
Misinformation hashtags such as `CONSPIRACY' ($k=1,056$), `FAKENEWS' ($k=6,686$), `TRUTH' are identified as well. 
China-related conspiracy theory hashtags including `CHINAVIRUS'($k=3,273$), 'CHINESEVIRUS` ($k=2,795$), `WUHANVIRUS' ($k=2,757$) and human rights hashtags such as `BLACKLIVESMATTER' ($k=6,390$) and `METOO' ($k=1,073$) existed in the class as well.
\subsubsection{class 2, $n=93$, purple: News}
The conspiracy-theory related hashtags, `WHO' ($k=8,042$) and `5G'($k=3,364$) are spotted in the class. In addition, science-related hashtags such as `VACCINES' ($k=10,843$), `SCIENCE' ($k=4,698$), `RESEARCH' ($k=1,459$),`HEALTHCARE' ($k=3,185$) and `CLIMATECHANGE' ($k=3,123$) are spotted. 
\subsubsection{class 6, $n=73$, cyan: Daily life}
This class comprises people's daily life amid the pandemic including `STAYHOME' ($k=17,960$), `SOCIALDISTANCING' ($k=12,957$) and 'QUARANTINE($k=12,623$)'. Meanwhile, we identified religious hashtags, including `GOD' ($k=1,112$) and `JESUS' ($k=1,953$). Top 40 degree hashtags of modularity class 0, 2, 6 are shown in Table \ref{tab: QAnon_neighbor}. 

\begin{table}[H]
\caption{Top 20 popular hashtags in class 0, 2, and 6 of QAnon hashtag co-occurrence network.}
\label{tab: QAnon_neighbor}
\centering\scriptsize
\begin{tabulary}{\linewidth}{@{}cLLL@{}}\toprule 
Rank & \centering 0 & \centering 2 & \centering\arraybackslash 6 \\
\cmidrule(r){1-1} 
\cmidrule(rl){2-2} 
\cmidrule(rl){3-3} 
\cmidrule(rl){4-4} 
1  & TRUMP            & PANDEMIC     & STAYHOME          \\
2  & USA              & CHINA        & QUARANTINE        \\
3  & COVIDIOTS        & VACCINE      & SOCIALDISTANCING  \\
4  & WEARAMASK        & WHO          & STAYATHOME        \\
5  & FAKENEWS         & NEWS         & MASKS             \\
6  & BLACKLIVESMATTER & HEALTH       & MASK              \\
7  & AMERICA          & US           & NYC               \\
8  & DONALDTRUMP      & UK           & QUARANTINELIFE    \\
9  & WUHAN            & VACCINES     & TWITTER           \\
10 & MAGA             & CANADA       & TIKTOK            \\
11 & FLORIDA          & ECONOMY      & STAYHOMESAVELIVES \\
12 & NEWYORK          & HEALTHCARE   & FACEMASK          \\
13 & CDC              & SCIENCE      & LOVE              \\
14 & BIDEN            & CALIFORNIA   & TRENDING          \\
15 & COVIDIOT         & COVIDVACCINE & YOUTUBE           \\
16 & TEXAS            & 5G           & MEMES             \\
17 & BLM              & COVID19UK    & CORONAPOCALYPSE   \\
18 & CORONAVIRUSUSA   & MEDIA        & THURSDAYTHOUGHTS  \\
19 & CNN              & FACEMASKS    & FRIDAYTHOUGHTS    \\
20 & BILLGATES        & AUSTRALIA    & LOCKDOWN2020      \\ 
   \bottomrule 
\end{tabulary} 
\end{table}

\begin{figure*}[t!]
    \centering
    \begin{subfigure}[b]{0.475\linewidth}
        \centering
        \includegraphics[width=\linewidth]{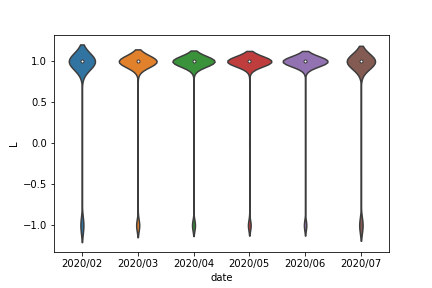}
        \caption{}\label{fig:violin1}
    \end{subfigure}%
    \hfill
    \centering
    \begin{subfigure}[b]{0.475\linewidth}
        \centering
        \includegraphics[width=\linewidth]{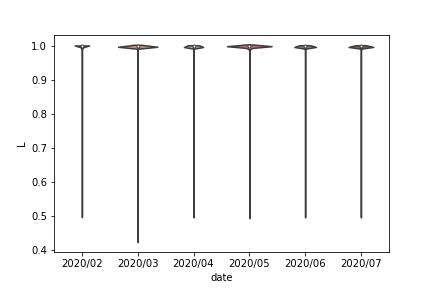}
        \caption{}\label{fig:violin2}
    \end{subfigure}%
    \hfill
    \begin{subfigure}[b]{0.475\linewidth}  
        \centering 
        \includegraphics[width=\linewidth]{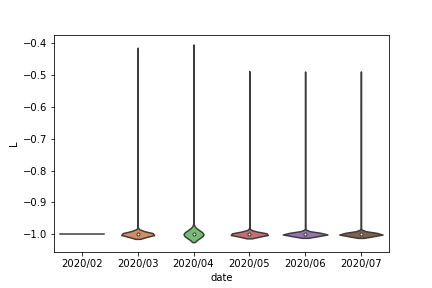}
        \caption{}\label{fig:violin3}
    \end{subfigure}
    \hfill
\caption{Distributions of QAnon-leaning ($L$) for (a) all users, (b) pro-leaning users, and (c) anti-leaning users. $L$ for less-leaning users is shown in Fig.~\ref{fig:violin} }  
\label{fig:violinplots}
\end{figure*}


\begin{sidewaystable}[pt!]
\caption{Topics extracted using LDA}  
\small
\begin{tabular}{|c|l|}
\toprule
Topic ID &
  \multicolumn{1}{c|}{Topics} \\ \hline
1 &
  0.017*"trump" + 0.011*"health" + 0.009*"president" + 0.008*"national" + 0.007*"public" + 0.006*"guard" + 0.004*"house" + 0.004*"news" + 0.003*"world" + 0.003*"former" \\ \hline
2 &
  0.010*"news" + 0.010*"april" + 0.008*"health" + 0.006*"vitamin" + 0.005*"national" + 0.005*"york" + 0.004*"trump" + 0.004*"last" + 0.004*"fox" + 0.004*"president" \\ \hline
3 &
  0.010*"health" + 0.008*"trump" + 0.006*"chinese" + 0.006*"news" + 0.005*"president" + 0.005*"last" + 0.004*"government" + 0.004*"public" + 0.004*"china" + 0.004*"world" \\ \hline
4 &
  0.013*"april" + 0.008*"march" + 0.007*"news" + 0.005*"member" + 0.005*"last" + 0.005*"health" + 0.005*"york" + 0.004*"december" + 0.004*"january" + 0.004*"first" \\ \hline
5 &
  0.019*"health" + 0.011*"house" + 0.010*"national" + 0.007*"world" + 0.007*"committee" + 0.006*"public" + 0.006*"congress" + 0.006*"president" + 0.006*"law" + 0.006*"york" \\ \hline
6 &
  0.018*"health" + 0.009*"news" + 0.007*"trump" + 0.006*"president" + 0.005*"house" + 0.005*"april" + 0.005*"county" + 0.005*"public" + 0.004*"government" + 0.004*"rate" \\ \hline
7 &
  0.007*"york" + 0.006*"health" + 0.005*"april" + 0.005*"death" + 0.004*"rate" + 0.004*"trump" + 0.004*"president" + 0.004*"world" + 0.004*"news" + 0.004*"last" \\ \hline
8 &
  0.005*"health" + 0.005*"april" + 0.004*"vitamin" + 0.004*"president" + 0.004*"york" + 0.003*"public" + 0.003*"news" + 0.003*"house" + 0.003*"trump" + 0.003*"first" \\ \hline
9 &
  0.009*"house" + 0.007*"news" + 0.005*"care" + 0.004*"trump" + 0.004*"last" + 0.004*"york" + 0.004*"congress" + 0.004*"health" + 0.004*"april" + 0.004*"president" \\ \hline
10 &
  0.007*"news" + 0.006*"april" + 0.004*"health" + 0.004*"wuhan" + 0.004*"first" + 0.003*"fox" + 0.003*"chinese" + 0.003*"president" + 0.003*"march" + 0.002*"mask" \\ \hline
11 &
  0.011*"health" + 0.007*"york" + 0.006*"house" + 0.005*"president" + 0.005*"trump" + 0.005*"city" + 0.004*"government" + 0.003*"former" + 0.003*"national" + 0.003*"world" \\ \hline
12 &
  0.012*"april" + 0.006*"member" + 0.006*"york" + 0.006*"health" + 0.006*"home" + 0.006*"national" + 0.005*"nursing" + 0.004*"president" + 0.004*"scholar" + 0.004*"news" \\ \hline \bottomrule
\end{tabular}
\label{tab:LDA}
Coefficient values indicate the importance of each word in a topic.
\end{sidewaystable}

\begin{figure}[pt!]
  \includegraphics[width=\linewidth]{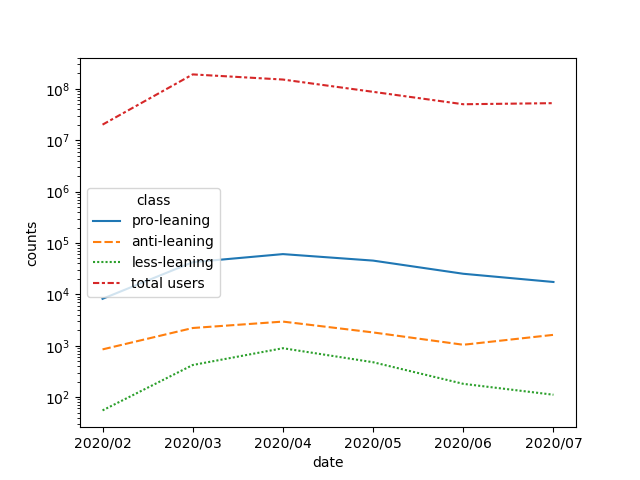}
\caption{The number of retweets for pro-leaning, anti-leaning, and less-leaning users and total.}
\label{fig:dynamic_retweets}
\medskip
\end{figure}
\end{appendices}
\end{document}